\documentclass[journal,comsoc]{IEEEtran}
\usepackage[T1]{fontenc}
\usepackage{amsmath}
\usepackage[cmintegrals]{newtxmath}
\usepackage{url}
\usepackage{soul,color,graphicx,float,epstopdf}
\usepackage{tablefootnote,textcomp,gensymb}
\usepackage{eurosym,booktabs,multirow}
\usepackage{amsfonts} 
\usepackage{algorithmic}
\usepackage{array}
\usepackage{stfloats}
\usepackage{float}
\usepackage{mathtools}
\usepackage{graphicx}
\usepackage{pdfpages}
\usepackage{comment}
\usepackage{balance}
\usepackage{xcolor}
\usepackage{comment,enumitem}
\usepackage[normalem]{ulem}
\usepackage[implicit=false]{hyperref}
\usepackage{subfigure}

\begin{document}

\title{Towards Practical Operation of Deep \\Reinforcement Learning Agents in Real-World Network Management at Open RAN Edges}

\author{Haiyuan Li,
Hari Madhukumar,
Peizheng Li,
Yuelin Liu,
Yiran Teng,\\
Yulei Wu,
Ning Wang,
Shuangyi Yan, 
Dimitra Simeonidou

\thanks{H. Li, H. Madhukumar, Y. Liu, Y. Teng, Y. Wu, N. Wang, S. Yan, and D. Simeonidou are with the High-Performance Networks Group, Smart Internet Lab, Department of Electrical and Electronic Engineering, University of Bristol, BS8 1QU, U.K. (e-mail: ocean.h.li@bristol.ac.uk).}

\thanks{P. Li is with Bristol Research and Innovation Laboratory, Toshiba Europe Ltd., BS1 4ND, U.K.}
}

\markboth{\LaTeX\ Class Files,~Vol.~X, No.~Y, Month~202X}%
{Shell \MakeLowercase{\textit{ - et al.}}: Bare Demo of IEEEtran.cls for Journals}

\maketitle 

\begin{abstract}
Deep Reinforcement Learning (DRL) has emerged as a powerful solution for meeting the growing demands for connectivity, reliability, low latency and operational efficiency in advanced networks. However, most research has focused on theoretical analysis and simulations, with limited investigation into real-world deployment. To bridge the gap and support practical DRL deployment for network management, we first present an orchestration framework that integrates ETSI Multi-access Edge Computing (MEC) with Open RAN, enabling seamless adoption of DRL-based strategies across different time scales while enhancing agent lifecycle management. 
We then identify three critical challenges hindering DRL’s real-world deployment, including (1) asynchronous requests from unpredictable or bursty traffic, (2) adaptability and generalization across heterogeneous topologies and evolving service demands, and (3) prolonged convergence and service interruptions due to exploration in live operational environments. To address these challenges, we propose a three-fold solution strategy: (a) advanced time-series integration for handling asynchronized traffic, (b) flexible architecture design such as multi-agent DRL and incremental learning to support heterogeneous scenarios, and (c) simulation-driven deployment with transfer learning to reduce convergence time and service disruptions. Lastly, the feasibility of the MEC-O-RAN architecture is validated on an urban-wide testing infrastructure, and two real-world use cases are presented, showcasing the three identified challenges and demonstrating the effectiveness of the proposed solutions.
\end{abstract}


\section{Introduction}
\label{sec:introduction}
Emerging frameworks for future networks, exemplified by ITU-R’s IMT-2030 initiative \cite{series2023imt}, outline a new generation of usage scenarios. Building upon the three scenarios recognized in IMT-2020, immersive communication, massive communication, and hyper-reliable/low-latency communication, IMT-2030 added three new categories: ubiquitous connectivity, service-oriented AI and communication, and integrated sensing and communication. These scenarios envision a future network that can deliver better connectivity, lower latency, and higher reliability for an increasing number of users with escalating demands on radio and computational resources~\cite{li2024cloud}.
Achieving these objectives necessitates two fundamental measures: (i) the implementation of more robust network architectures to effectively address increasingly dynamic and diverse challenges, and (ii) the adoption of efficient management strategies for the network’s capabilities.

In terms of network orchestration, standardized approaches like ETSI Multi-Access Edge Computing (MEC)~\cite{ETSIMEC} and Open RAN (O-RAN)~\cite{ORAN-WG1} are particularly advantageous. MEC provides a common orchestration framework and APIs that enable on-demand deployment and lifecycle management of MEC applications. This not only ensures fine-grained control over allocated computing resources but also guarantees real-time network management. On the other hand, O-RAN disaggregates traditional Radio Access Network (RAN) functions into different modules, offering open interfaces such as A1, E2, and O1. These open and well-defined APIs permit vendor-agnostic integration of intelligence-driven functionalities. By unifying MEC’s orchestration features with O-RAN’s modular RAN functions, operators can flexibly deploy, manage, and scale intelligence-driven functionalities across the network.

Furthermore, in the context of improving network management efficiency, Deep Reinforcement Learning (DRL) is widely recognized as a key enabler in this effort~\cite{alwarafy2022frontiers}. Compared to conventional approaches that rely on domain-specific rules, formal optimization, or data-driven learning from labeled or unlabeled data, DRL gains a distinct advantage by tying reward functions to service-level objectives and captures both immediate outcomes and long-term impacts through the optimization of cumulative rewards~\cite{hurtado2022deep}. However, to satisfy the aforementioned network demands, relying on model-based offline learning solutions proves insufficient, as explicit or approximate representations of network dynamics and reliance on historical datasets may fail to capture the complexity and diversity of real-world conditions~\cite{li2020accelerating, ye2020model}. These shortcomings become pronounced under increasing demands for throughput, latency, and mobility and can result in suboptimal network performance. Consequently, model-free online learning solutions that directly interact with real-world networks, rather than depending on a predefined model, are more adaptable in highly dynamic communication scenarios~\cite{ginzburg2024reinforcement}.

However, although extensive theoretical research has explored model-free online learning solutions, such as Policy- and Value-based, in network management, deploying these DRL algorithms in real-world networks remains a pressing challenge. In response, this article undertakes a comprehensive investigation to set the stage for practical DRL deployment in O-RAN Edges, and the main contributions are summarized as follows:

\begin{figure*}[t]
    \centering
    \includegraphics[width=1\linewidth]{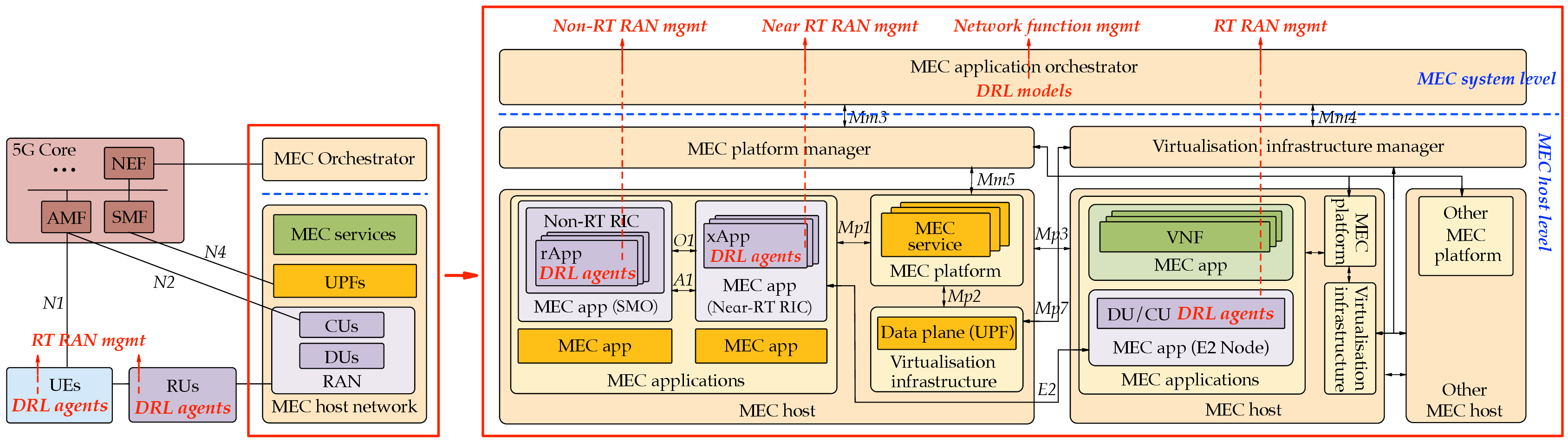}
    \caption{A future network architecture incorporating RAN, MEC, and 5G Core functions based on frameworks proposed by ETSI~\cite{ETSIMEC} and O-RAN Alliance~\cite{ORAN-WG1}.}
    \label{fig:architecture}
\end{figure*}

\begin{itemize}[leftmargin=*]
\item \textit{Unified DRL Deployment Architecture.} In bringing DRL solutions into real-world networks, a key consideration involves creating an architecture that accommodates diverse DRL agents, varying timescales, and distinct training and inference requirements. To address these demands, we introduce a unified framework built on foundational principles stemming from O-RAN and MEC, ensuring consistent interfaces for state collection, action execution, and reward evaluation. By leveraging the modular architecture of O-RAN and the orchestration capabilities of MEC, this framework seamlessly integrates DRL-driven processes across Real-Time, Near/Non-Real-Time timescales.
It also ensures that the placement of DRL agents, along with the partitioning of training and inference tasks, can be flexibly tailored to specific applications.

\item \textit{Identification of Three Key Challenges in Real-Life Network Operations.}
While the integrated MEC–O-RAN framework lays a solid foundation for DRL-based network management, real-world conditions introduce further complexities beyond theoretical analysis. We identify three primary obstacles of DRL in real-world deployment: \emph{(i) asynchronised requests}, which arise from irregular or bursty state transitions and complicate the assumption of a stationary environment; \emph{(ii) adaptability and generalization}, as learned policies often fail to transfer across heterogeneous topologies, dynamic traffic patterns, and evolving service requirements; and \emph{(iii) service interruptions and extended convergence}, because safe exploration and fast learning are particularly difficult to achieve in live operational networks with strict Quality of Service (QoS) constraints. 

\item \textit{Mitigation Solutions.} In response, we propose a three-pronged solution for enabling practical DRL-based network management: \emph{(i) advanced time-series integration}, which incorporates forecasting models into the learning process to handle fluctuating traffic arrivals; \emph{(ii) flexible architecture design}, including decoupled information profiling, multi-agent DRL, and incremental learning, to ensure adaptability across heterogeneous topologies and evolving service requirements; and \emph{(iii) safe exploration mechanisms}, leveraging simulation-based pretraining, transfer learning, and constraints on agent behavior to accelerate convergence and minimize service disruptions under strict QoS constraints.

\item \textit{Real-World Validation.} Lastly, We validate the proposed MEC-O-RAN orchestration on an urban-wide testing infrastructure built with OpenStack, Kubernetes, and DevOps tools. Two representative real-world trials, task offloading and network slicing, are then implemented, illustrating how the identified challenges arise in diverse networking scenarios and confirming the effectiveness of our solutions in mitigating these challenges.
\end{itemize}

\section{Network orchestration for DRL applications}
\label{sec:orchestration}

In practical DRL-based network management applications, a fundamental concern arises from determining the optimal venue of DRL agents' placement across the network. Essentially, it requires a flexible orchestrator that i) accommodates various timescales ranging from milliseconds to seconds, ii) manages the training and inference phases of DRL, 
and iii) provides consistent interfaces for state collection and action execution.

Therefore, we introduce a unified framework as shown in Fig.~\ref{fig:architecture} that integrates the ETSI MEC with the O-RAN architecture, offering a comprehensive solution to these deployment concerns. In this architecture, 
the O-RAN components, including Distributed Unit (DU), Centralized Unit (CU), Near-Real-Time (RT) RAN Intelligent Controller (RIC), Non-RT RIC, and Service Management and Orchestration (SMO), as shown in purple in Fig.~\ref{fig:architecture}, can be implemented as containerized MEC applications. Furthermore, rApps (in the Non-RT RIC), and xApps (in the Near-RT RIC) can operate on a common MEC infrastructure that oversees deployment, lifecycle management, and resource allocation. 
Moreover, it is worth noticing that this containerized MEC framework is not limited to RAN functions. The same framework environment can host DRL-based solutions for edge computing, content caching, and traffic steering, allowing operators to deploy diverse applications alongside communication services. 

DRL agents can be deployed among these network components, spanning three different timescales.
At the RT level, DRL agents can be deployed directly on the UE or integrated within distributed DU/CU functions to dynamically adjust real-time parameters, such as transmit power or handover triggers, on a millisecond scale.
At the Near-RT level, DRL modules can be positioned into xApps to manage tasks such as scheduling, interference management, and load balancing over timescales ranging from 10 milliseconds to 1 second.
At the Non-RT level, DRL agents can be implemented in rApps and the SMO to oversee large-scale resource provisioning, network slicing, and other strategic optimizations over longer intervals, typically on the order of seconds or more.
By distributing DRL agents across these time scales, operators can achieve a balance between localized, short-term adaptability at the radio edge and global, long-term policy enforcement across the broader network.

Beyond time-scale distribution, this orchestration can separate training and inference stages while dynamically allocating resources as needed.
For instance, MEC orchestrators can allocate additional CPU/GPU resources to rApps or the SMO for training workloads. Once training is completed, the SMO or rApps deploy the refined policy to xApps or CUs/DUs for near-real-time decision-making, freeing up resources for other computing demands.
Lastly, leveraging the unified orchestration capabilities, these services can coexist seamlessly while meeting their individual performance requirements. By harmonizing O-RAN interfaces (A1, E2, O1) with MEC APIs (Mp1, Mp2, Mp3), the framework provides consistent interfaces for state collection and action execution, ensuring that DRL agents can incorporate both RAN-level telemetry (e.g., cell load, UE mobility) and edge service metrics (e.g., server utilization, content popularity) into their decision-making processes.


Overall, this integrated MEC and O-RAN approach forms a cohesive solution for integrating DRL into live 5G networks. The RAN stack, from UE, RU, DU, and CU-UP to the RICs and SMO, runs atop virtualized resources administered by MEC, facilitating dynamic scaling for both training and inference tasks. By embedding other MEC-based services alongside RAN functionalities, the proposed architecture facilitates end-to-end intelligence spanning communication and computing domains. Consequently, the architecture provides a scalable platform wherein DRL can operate at multiple timescales, powering advanced applications that enhance network efficiency, user experience, and operational agility.

\section{Challenges toward DRL's practical viability}
\label{sec:challenges}
While the unified framework introduced in Section~\ref{sec:orchestration} lays the groundwork for deploying DRL in real networks, as various DRL agents embedded in different network components interact with the actual network environment, a range of challenges may arise due to the inherent complexities of real-world networks that are often overlooked in simulation-based scenarios. In this section, we present three critical challenges and discuss the difficulties these challenges pose for DRL in practical deployments.

\subsection{Asynchronous requests}
DRL operates through a sequence of discrete decision steps defined by a Markov Decision Process (MDP). In network management, these steps are often triggered by traffic arrivals, and a key challenge arises from the variability in the request arrival frequencies.
In simulation environments, this aspect is often overlooked, with the state assumed to progress at fixed intervals. However, in real-world scenarios, event timings may fluctuate significantly, leading to non-uniform state transitions that can degrade learned policy performance. 

For instance, in resource-constrained environments, back-to-back task arrivals can amplify the impact of prior decisions while prolonged idle periods may undermine the effectiveness of previously cautious resource-saving strategies. This variability in the state space, which evolves across multiple time scales, complicates state transition modeling and credit assignment and will ultimately reduce the effectiveness of DRL.

\subsection{Adaptability and generalization}

Another critical challenge involves extending pre-optimized DRL models to diverse and evolving 5G networks. A DRL model often relies on a neural network tailored to a specific input–output structure and operating scenario. As a result, a policy trained in a fixed network context may not generalize to another with different configurations that alter the state and action dimensions. Consequently, deploying a single DRL solution across heterogeneous networks remains challenging without retraining or redesigning the model.

Beyond structural differences, dynamic changes and unexpected scenarios in network traffic and demand further complicate the adaptability of DRL agents. In 5G networks, load levels fluctuate, user behavior evolves, and new services or failures can abruptly modify the operating status. 
Although a DRL agent may excel under the specific conditions for which it was trained, it struggles when presented with scenarios that extend beyond its learned experience.
This underscores a critical gap in adaptability, where the agent must not only generalize to unseen states but also adjust its behavior as the environment evolves.

\subsection{Extended convergence and service interruptions} 
The prolonged convergence times of DRL in live networks present a significant operational challenge. Unlike simulation environments, where the experiment process can be accelerated to expedite training, real-world networks operate in time sequence without shortcuts. In practice, achieving a well-optimized DRL policy often needs extensive exploration and thorough hyperparameter tuning, making the training process highly time-consuming. Moreover, frequent trial-and-error exploration can cause unnecessary operations and disruptive actions that degrade performance and violate Quality-of-Service (QoS) guarantees.

\section{Mitigation solution strategies of these challenges}
\label{sec:solutions}
To address the aforementioned challenges, we introduce three specialized solutions to be called on demand. 
Predictive time-series modeling provides DRL foresight capability and strengthens its long-term planning ability to optimize cumulative rewards.
Network extraction with incremental learning focuses on handling heterogeneous and continuously shifting network scenarios. 
Lastly, simulation-driven training with transfer learning allows DRL for thorough exploration and refinement in a controlled setting.
These solutions can be integrated to offer a cohesive and adaptive framework that scales effectively and preserves service quality.


\subsection{Predictive time-series integration}
In dealing with the challenge posed by non-uniform and asynchronous request arrivals in networks, an effective strategy is to integrate time-series prediction models, such as Long Short-Term Memory (LSTM) networks or Transformers, into the DRL framework. 
Initially, these offline models can be pre-trained on historical request data to capture the hidden temporal patterns (e.g., daily or weekly traffic cycles) and fluctuations in arrival rates (e.g., sudden bursts of requests).
Once the predictor is ready, it can serve as a front-end module for online DRL. Rather than relying exclusively on the original state information, it can incorporate forecasts of future arrival times and related details derived from historical data as additional inputs. By anticipating upcoming requests, the agent can avoid over-committing resources during periods of high demand, while also preventing excessive conservatism that would otherwise lead to idle resources during periods of minimal or nonexistent demand. Furthermore, by forecasting future trends, the predictive model enables the DRL agent to adjust its management strategies in advance to ensure real-time service.

\subsection{Network extraction and incremental learning}
A practical strategy to address dynamic and heterogeneous network conditions is to decouple information profiling from the DRL pipeline. In this approach, raw and diverse network data are first normalized or translated into a uniform representation, such as feature embeddings or standardized tensors, before they are fed into the DRL model~\cite{li2025achieving}. This approach eliminates the impact of changes in traffic patterns, device configurations, or network slice parameters on input dimensions, therefore, enhancing the generalization of DRL. 
To further improve flexibility, multi-agent DRL can be employed. Instead of relying on a single monolithic agent that handles all network functionalities, multiple agents can be deployed and scaled in accordance with different slices, services, or functionalities. This granular distribution of control reduces complexity, facilitates parallel training, and improves fault tolerance. Each agent specializes in its designated domain while collectively adapting to broader network dynamics.
Finally, to cope with drifting or evolving network scenarios, techniques such as incremental learning can be adopted. Rather than retraining a DRL policy from scratch whenever conditions change, an incremental approach updates the existing policy with newly observed data and conditions. This approach maintains an up-to-date model with minimal overhead and delivers rapid adaptation to unexpected network variations. By combining data decoupling, multi-agent architectures, and incremental learning, DRL solutions can be more robust and responsive to the diverse and evolving nature of 5G networks.


\subsection{Simulation-driven deployment with transfer learning}
A practical strategy to reduce training overhead and mitigate operational risks is to train DRL models in a controlled environment and then adapt them to real-world conditions. In a controlled setting, extensive exploration does not threaten network performance or QoS guarantees, and accelerated simulations can expedite hyperparameter tuning. To enhance the robustness of the algorithm, domain randomization or domain adaptation can expose the DRL policy to diverse simulated conditions, preparing it for unexpected scenarios in real-world networks. Once a policy demonstrates promising results in simulation, transfer learning methods can help bridge the gap to a live network by fine-tuning the policy with fewer real-world interactions. Safety constraints or QoS-focused objectives can be incorporated through constrained RL methods or reward shaping, penalizing unacceptable performance and discouraging undesirable actions. 
To ensure the feasibility of transfer learning, two key requirements must be guaranteed: (1) high-fidelity simulation to accurately capture real-world complexity, and (2) a deployment venue in the real network that provides access to the same state information and action space as those available in the simulation~\cite{li2022sim2real}.

\section{Testing infrastructure implementation and use cases}
\label{sec:testbed}
The previous sections examine the MEC-O-RAN architecture that supports DRL deployment, the key challenges in practical applications, and the corresponding mitigation solutions. To validate these discussions, we first construct an experimental platform based on the MEC-O-RAN architecture presented in Fig.~\ref{fig:architecture}.
Based on this infrastructure, we then implement two distinct use cases: task offloading and network slicing, where task offloading focuses on the first challenge of event frequency variation, while network slicing showcases the latter two challenges and demonstrates how multiple issues can emerge in a single setting. Lastly, we deploy the proposed solutions and demonstrate their broad effectiveness across different contexts.

\begin{figure}[b]
    \centering
    \includegraphics[width=1\linewidth]{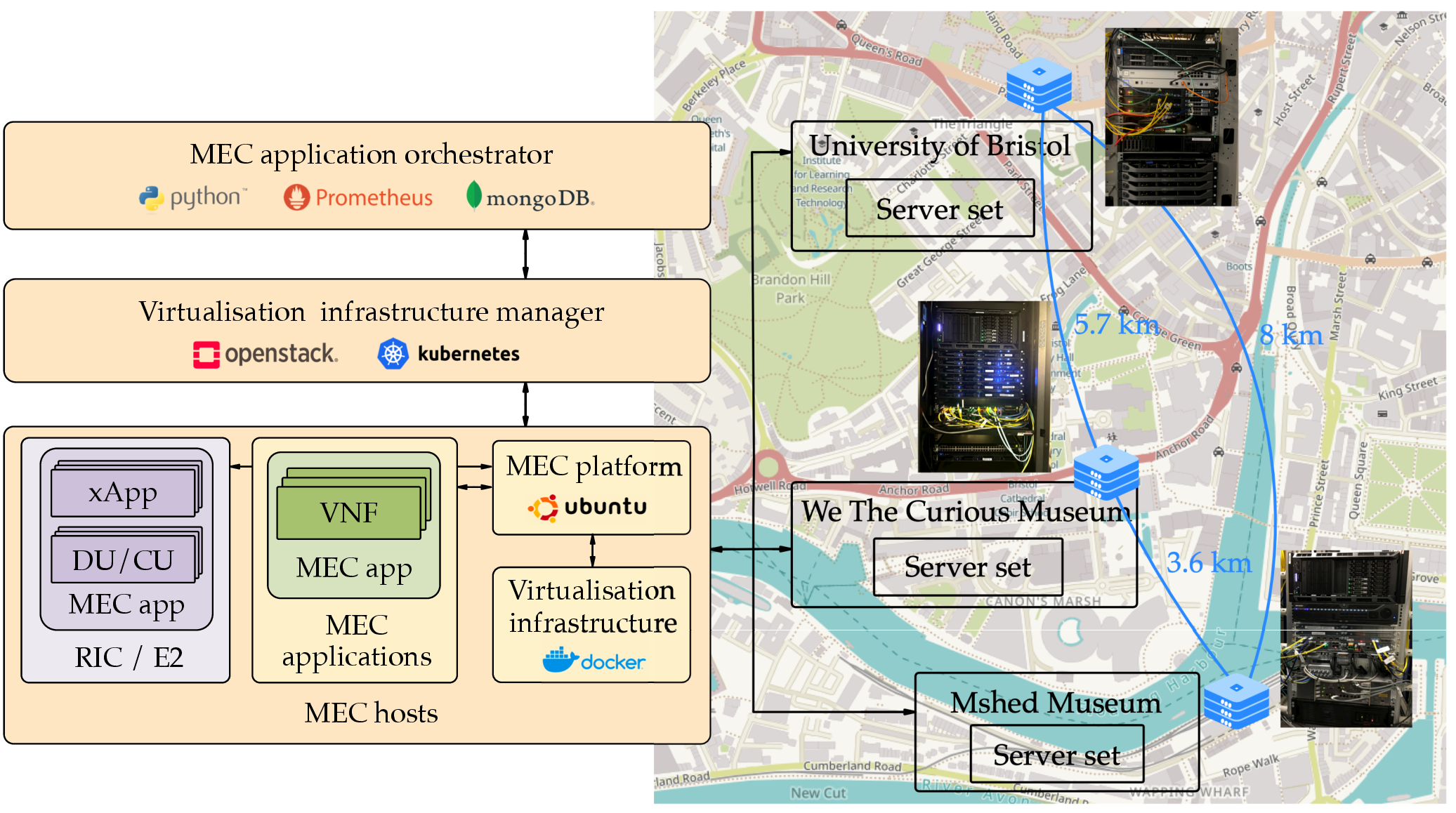}
    \caption{An urban network with three distributed MECs based on OpenStack and Kubernetes.}
    \label{fig:testbed}
\end{figure}

\subsection{Testing platform deployment and DRL application pipeline}
The practical implementation of the orchestration (as shown in Fig.~\ref{fig:architecture}) on the testing infrastructure is shown in Fig.~\ref{fig:testbed}.
From the bottom layer upward, we provision three geographically separated physical servers, located at the University of Bristol, the We the Curious Museum, and the Mshed Museum, to serve as MEC hosts. Each MEC host runs on an Ubuntu-based platform with Docker installed to support container-based virtualization. This setup enables flexible deployment of lightweight MEC applications and VNFs at the edge. 
Above this host layer, we employ a Virtualization Infrastructure Manager (VIM) realized through a combination of OpenStack and Kubernetes~\cite{kaur2022review}. OpenStack is utilized to create Virtual Machines (VMs), each of which serves as a MEC node. A Kubernetes cluster is then deployed across these VMs, where network functions and task-processing modules are encapsulated as pods. CPU resources for each pod are controlled via Kubernetes resource limits and stressed using the \texttt{stress} command, while link capacity is managed through the Linux \texttt{tc} utility and saturated with \texttt{iperf}. 
At the orchestration layer, network management components, such as the MEC application orchestrator, xApps, and rApps, can utilize these \textit{action interfaces} to communicate with the Kubernetes APIs to allocate network resources in response to DRL decisions. 
To obtain the \textit{state and rewards}, each node periodically collects performance metrics, such as CPU utilization and network throughput, using Prometheus \texttt{NodeExporter} agents deployed in a \texttt{DaemonSet} configuration. These time-stamped metrics are stored in a MongoDB database, enabling the DRL agent or orchestrator to retrieve up-to-date network status. 
In this network, the communications between various network components are mediated by FastAPI-based services, and the design is readily adaptable to standard ETSI MEC and O-RAN interfaces.

\begin{figure}[t] 
    \centering  
            \includegraphics[width=0.93\linewidth]{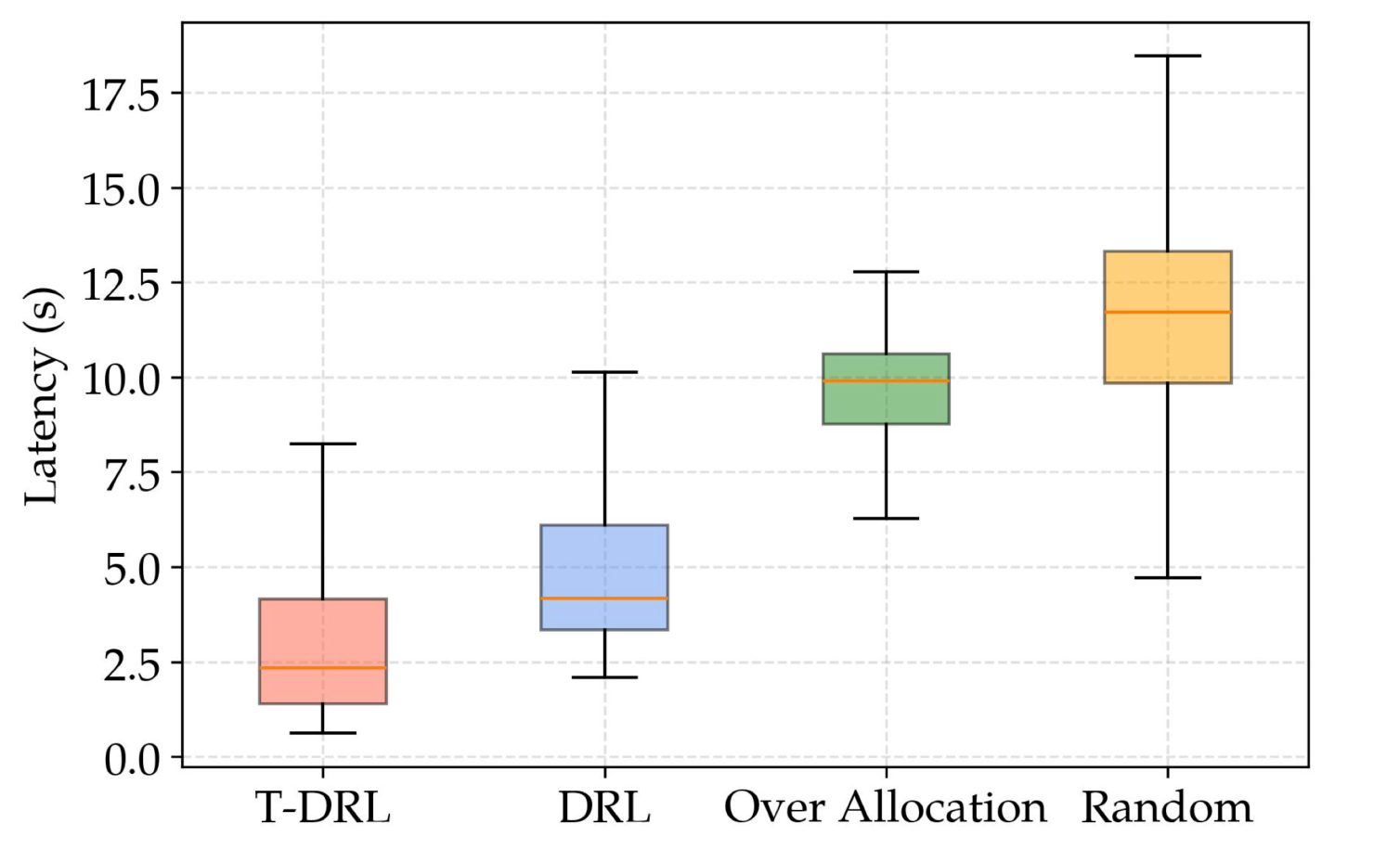}
    \caption{Performance evaluation of the Transformer-assisted DRL with synchronized requests.}
    \label{fig:result0}
\end{figure}

\subsection{Use case I: task offloading}
\subsubsection{Asynchronized request challenge}
We first consider a task offloading scenario in an eight-node MEC network, where each node is provisioned with 4 CPU cores and constrained by a 10 Gbps inter-node bandwidth limit. Offloading requests are emulated using the Google Cluster Traces dataset~\cite{reiss2011google}, which provides actual arrival times and computational resource demands. Asynchronized incoming requests distinguish our setup from many simulation-based studies that rely on fixed, discrete time steps. It poses the first challenge of event frequency variation, where the rewards of resource allocation decisions are influenced by future request patterns. To address this issue, we implemented the proposed solution in Sec.~\ref{sec:solutions} by augmenting our DRL mechanism with transformer-based predictions~\cite{ramana2023vision} for both task arrival times and resource needs. The Transformer component can anticipate near-future workload fluctuations by analyzing historical request patterns, enabling the DRL to allocate resources proactively.

\subsubsection{Experiment results}
We compare the latency distribution across four approaches: (1) T-DRL, our DRL-assisted method augmented by a transformer-based predictor; (2) DRL, a conventional RL solution without predictive modeling; (3) Over-Allocation, a static approach that reserves extra resources; and (4) Random, which arbitrarily assigns resources. As shown in Fig.~\ref{fig:result0}, the Over-Allocation strategy leads to inefficient resource usage and increases latency when multiple tasks contend for the shared resources. 
Likewise, the Random method suffers from unpredictable assignment patterns, resulting in the highest latency range. 
In contrast, the DRL-based solution with long-term planning capability can effectively reduce computing latency compared to static approaches. 
Nevertheless, as can be seen in Fig.~\ref{fig:result0}, with the assistance of a traffic predictor, the proposed T-DRL strategy can further improve the performance by 35\% in terms of average latency, demonstrating the effectiveness of considering future traffic patterns in facilitating network management.



\begin{figure}[b]
    \centering
    \includegraphics[width=1\linewidth]{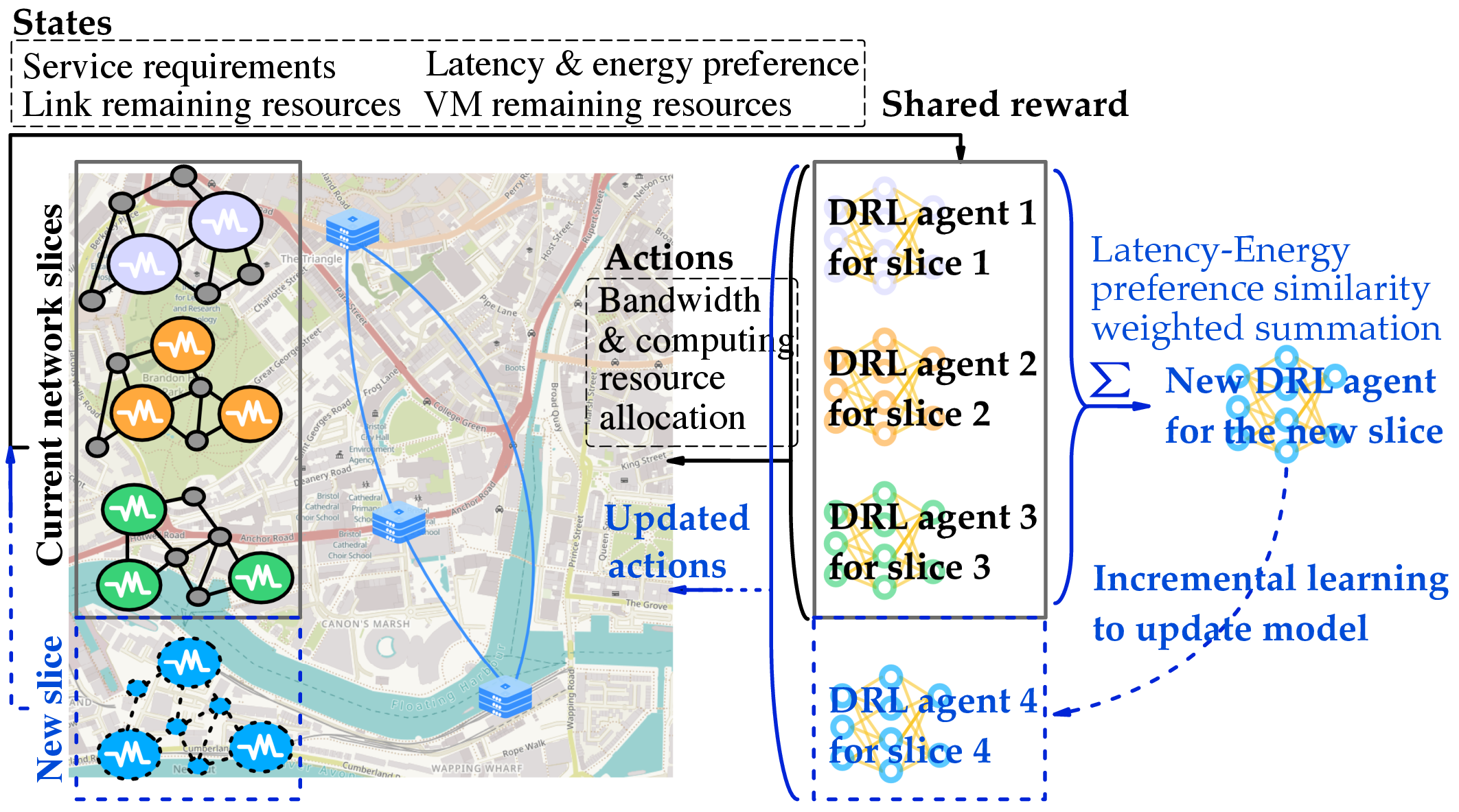}
    \caption{Overview of the proposed incremental learning, multi-agent DRL strategy for adaptive network slicing.}
    \label{fig:solution}
\end{figure}

\begin{figure}[t] 
    \centering  
    \subfigure[4-slice scenario]{
                    \includegraphics[width=0.9\linewidth]{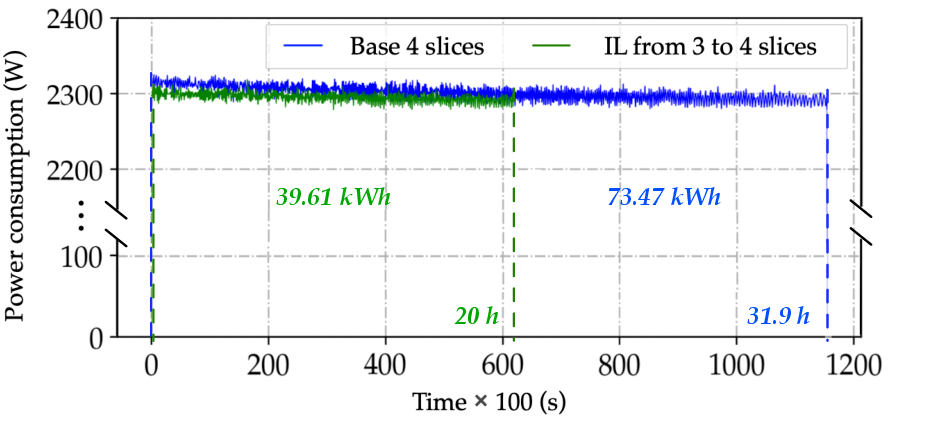}
                }
        \subfigure[5-slice scenario]{
                    \includegraphics[width=0.9\linewidth]{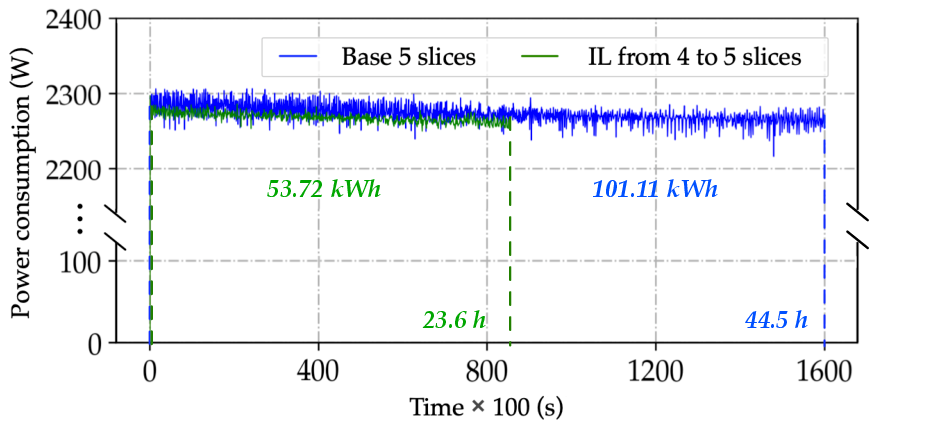}}
    \caption{Energy consumption comparison of multi-agent DRL between training from scratch strategy and incremental learning scenarios. The x-axis shows training time, and the y-axis indicates power consumption. The area under each curve represents the total energy consumed during training.}
	\label{fig:result1}
\end{figure}

\subsection{Use case II: network slicing}
\subsubsection{Adaptability and generalization challenge}
In the second use case, we examine a network-slicing scenario set in a six-node environment, employing the same resource configuration as in the previous example. Multiple slices co-exist to support diverse applications, and each slice spans several VNFs with distinct latency and energy requirements. Notably, each slice’s computational load and inter-VNF bandwidth demands range between 27\% and 33\% of the total available resources, introducing a high level of contention and necessitating flexible resource management.
While DRL-based methods have proven effective for resource allocation in slicing scenarios~\cite{hurtado2022deep}, the dynamic nature of slice deployment adds an extra layer of complexity. In this setting, slices may be instantiated or removed as services complete or reinitialize, altering the action space and disrupting established policies. The instantiation of the 'adaptability and generalization' necessitates continuous adaptation, often requiring DRL solutions to be updated to maintain effectiveness.

As discussed in Sec.~\ref{sec:solutions}, to accommodate these evolving resource demands without incurring prohibitive training overhead, we adopt a fully cooperative, multi-agent DRL architecture as shown in Fig.~\ref{fig:solution}. Each slice is managed by a dedicated agent, which helps isolate the effect of slice arrivals and departures. Additionally, we implement an incremental policy inheritance mechanism to minimize re-optimization when new slices appear. Specifically, the initial single-agent policy for a newly introduced slice is derived from a weighted average of existing agents’ parameters, where the weights reflect similarities in latency and energy demands between the new join slice and the existing ones. All agents then continue joint training under the updated slice composition to refine their management actions cooperatively.
When slices depart, their trained parameters can either be preserved for later reactivations or redistributed among remaining agents, thus avoiding the need for wholesale retraining.

\begin{figure}[!t] 
    \centering  
    \subfigure[Transfer learning from simulation]{
                    \includegraphics[width=0.9\linewidth]{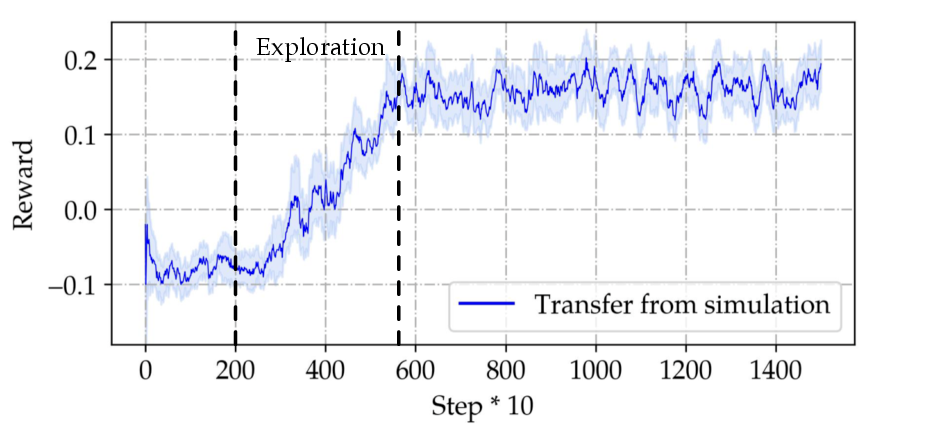}
                }
        \subfigure[Training from scratch on practical network]{
                    \includegraphics[width=0.9\linewidth]{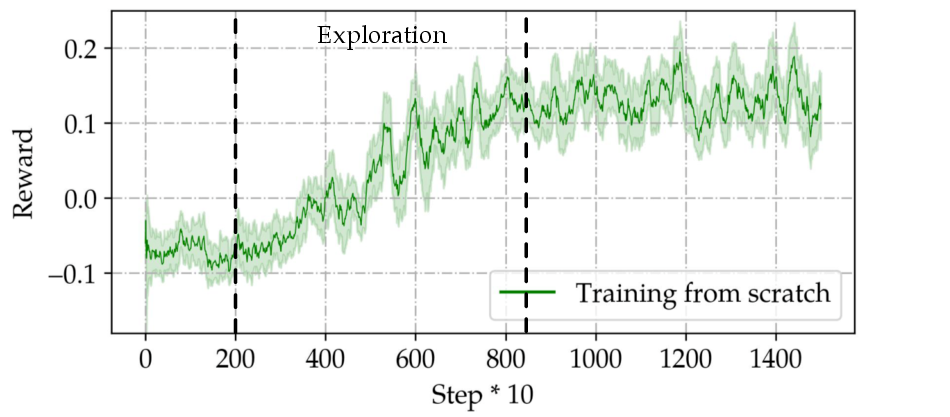}
                    }
    \caption{Comparison of DRL training approaches: Training from scratch in testbed vs. Transfer learning from a simulation-based pre-trained model.}
	\label{fig:result2}
\end{figure}

\subsubsection{Experiment results of incremental learning}
In Fig.~\ref{fig:result1}, we track the total energy consumption throughout the training process. Because training involves both the computational overhead of the DRL agent and the operational cost of the environment, we measure the entire experimental platform’s power usage. Remarkably, incremental learning demonstrates around a 50\% reduction in energy consumption and training time compared to a from-scratch approach. These results highlight the effectiveness of the proposed incremental learning solution in addressing the adaptability limitations of DRL, particularly in dynamic network environments that demand continuous efficiency optimization.

\subsubsection{Extended convergence and service interruptions challenge} 
The second challenge encountered in this setup is 'service interruptions and extended convergence'. In real-world deployments, function adjustments incur actual delays, which can significantly increase the time cost of large-scale interactions. As shown in Fig.~\ref{fig:result1}, training from scratch in a 5-slice scenario requires 44.5 hours.
To reduce the learning cost, we deployed the proposed simulation-driven deployment with transfer learning strategy. 
First, we construct a high-fidelity simulation environment with the same resource parameters, VNF configurations, and traffic demands as in the physical network. 
This offline setting allows us to train the DRL model under compressed timescales and accumulate experience without risking service interruptions. After the policy converges in the simulation environment, we deploy the model into the live network and perform only minimal fine-tuning to account for any discrepancies not captured in the simulator. In addition, this two-stage approach strikes a balance between exploration-driven learning and the need for service continuity, offering a practical framework for integrating DRL-based resource orchestration into dynamic, mission-critical operational networks.

\subsubsection{Experiment results of transfer learning}
Fig.~\ref{fig:result2} compares the reward trajectories for two training strategies: (a) transfer learning from a simulation-pretrained model and (b) training from scratch on the live network. The results indicate that while both approaches ultimately achieve similar reward levels, transfer learning exhibits a steeper ascent, suggesting faster convergence and 44\% of reduced exploration time compared to training from scratch, which progresses gradually and entails a higher risk of service interruptions. This outcome underscores the practical value of leveraging offline simulation to pre-train DRL agents before deployment in mission-critical Open RAN environments.

\section{Conclusion and Open Issues}
\label{sec:conclusion}
In this article, we examine practical pathways for deploying DRL agents in real-world networks. We begin by presenting a unified orchestration framework that incorporates ETSI MEC principles and O-RAN architecture, enabling seamless DRL integration across multiple network layers and time scales. Unlike simulation-based studies that overlook critical real-world complexities, this framework ensures that DRL applications can flexibly adapt to heterogeneous infrastructures and stringent latency requirements. We then identify three key deployment challenges that persist when transitioning from controlled simulations to live network environments and further propose corresponding potential mitigation solutions. 
Finally, an urban-wide testing infrastructure is implemented, which proves the feasibility of the designed architecture. Two use cases are deployed on this network, showcasing how the three identified challenges manifest in real-world applications and validating the effectiveness of the proposed solutions.

Nevertheless, large-scale DRL deployments may involve additional complexities, prompting further exploration of the following open issues. Below, we briefly outline two additional open challenges and potential pathways for addressing them:
\emph{i) Co-existence of multiple DRLs:} DRL can tackle a variety of network management tasks, which often intersect when multiple DRL agents operate on the same network. If each agent optimizes its own local objective, the overall system may experience unintended performance degradation. Conflicting resource allocations or timing mismatches can destabilize the network and hinder global efficiency. One line of research involves designing a hierarchical AI architecture with a top-level coordinator to monitor performance and orchestrate resolution strategies among subordinate DRL modules. Although early studies (e.g., Zhang \textit{et al.}~\cite{zhang2022team}) have shown promise by sharing action information among multiple agents, scaling this approach to tens or hundreds of AI modules remains an open challenge.
\emph{ii) Delayed reward \& retroactive influence of actions:} In practical deployments, rewards are often subject to communication or processing delays, which can obscure the immediate impact of each action. Moreover, subsequent actions can retroactively affect how previous actions are evaluated, further complicating the learning process. A promising strategy is to adopt a hybrid model-based, off-policy approach, which employs predictive modeling for interim reward estimates and a retrospective update mechanism to refine credit assignment once delayed feedback arrives. By allowing the agent to learn continuously despite latency, and by revisiting earlier trajectories to update credit when actual rewards become available, this approach can mitigate the uncertainty introduced by delayed signals and overlapping effects among sequential actions.

Overall, by offering both conceptual foundations and empirical insights, we hope this work can accelerate the practical adoption of DRL in network management, and encourage operators and researchers towards developing more intelligent, adaptive, and resilient communication systems.

\section*{Acknowledgement}
\small{The authors would like to express their gratitude for the support from the Future Telecoms Research Hub, Platform for Driving Ultimate Connectivity (TITAN), sponsored by the Engineering and Physical Sciences Research Council (EPSRC) under Grant EP/X04047X/2 and Grant EP/Y037243/1.
This work was also supported by UK-funded projects REASON under the Future Open Networks Research Challenge sponsored by the Department of Science, Innovation and Technology (DSIT).}

\bibliographystyle{IEEEtran} 

\bibliography{IEEEabrv,references}

\small{\noindent \textbf{Haiyuan Li} is a Research Associate at the University of Bristol (UoB).}

\small{\noindent \textbf{Hari Madhukumar} is a Research Associate and pursuing his Ph.D. at the UoB.}

\small{\noindent \textbf{Peizheng Li} is a Senior Research Engineer at the Bristol Research and Innovation Laboratory, Toshiba Research Europe Ltd.}

\small{\noindent \textbf{Yuelin Liu} is pursuing his Ph.D. at the UoB.}

\small{\noindent \textbf{Yiran Teng} is pursuing his Ph.D. at the UoB.}

\small{\noindent \textbf{Yulei Wu} is an Associate Professor at the UoB.}

\small{\noindent \textbf{Ning Wang} is a Professor at the UoB.}

\small{\noindent \textbf{Shuangyi Yan} is an Associate at the UoB.}

\small{\noindent \textbf{Dimitra Simeonidou} (FREng, FIEEE) is a Professor at the UoB. She is also a Fellow of the Royal Academy of Engineering, and a Fellow of the Institute of Electrical and Electronic Engineers.}

\end{document}